# *Generalized Box-Cox method to estimate sample mean and standard deviation for Meta-analysis*

Olivia Xiao[1], Stacy Wang[2] and Min Chen[3,*]

[1] Highland Park High School; 4220 Emerson Ave, Dallas, TX 75205
[2] Department of Economics, Massachusetts Institute of Technology; 77 Massachusetts Ave, Cambridge, MA 02139
[3] Department of Mathematical Sciences, the University of Texas at Dallas, 800 W Campbell Rd, Richardson, TX 75080
**\*** Correspondence: mchen@utdallas.edu

## ABSTRACT

*Meta-analysis is the aggregation of data from multiple studies to find patterns across a broad range relating to a particular subject. It is becoming increasingly useful to apply meta-analysis to summarize these studies being done across various fields. In meta-analysis, it is common to use the mean and standard deviation from each study to compare for analysis. While many studies reported mean and standard deviation for their summary statistics, some report other values including the minimum, maximum, median, and first and third quantiles. Often, the quantiles and median are reported when the data is skewed and does not follow a normal distribution. In order to correctly summarize the data and draw conclusions from multiple studies, it is necessary to estimate the mean and standard deviation from each study, considering variation and skewness within each study. In past literature, methods have been proposed to estimate the mean and standard deviation, but do not consider negative values. Data that include negative values are common and would increase the accuracy and impact of the me-ta-analysis. We propose a method that implements a generalized Box-Cox transformation to estimate the mean and standard deviation accounting for such negative values while maintaining similar accuracy.*

## INTRODUCTION

Meta-analysis is used in various fields, including medicine, finance, and education, to draw patterns across a wide range of studies. Due to the independent nature of most studies, they may report different summary statistics. The most common metric reported is mean and standard deviation. However, some studies report other summary statistics when the data is skewed. In order to analyze the data, it is important to have a consistent metric to compare across studies. In past literature, methods have been proposed to estimate the mean and standard deviation. Utilizing the Luo [1] and Wan [2] methods for calculating sample mean and standard deviation, respectively, McGrath et al.[3] formulated a new method to estimate mean and standard deviation. In using the Box-Cox (BC) transformation, the method cannot handle negative values. However, negative values are common in many types of data and are therefore beneficial to include when performing meta-analysis. When pooling data from multiple studies, only using positive results overshoots the significance of a finding and may be misleading. In the past, many studies have ignored negative results. However, it would be beneficial to include negative results in order to get a better idea of the whole effect of a study. According to Haidich [4], it is common to have divisive results from multiple independent studies that pertain to significant medical questions. Meta-analysis is used to summarize the findings in such studies to create a general evaluation of research on a certain topic and the key takeaways from this research. Meta-analysis is an important method in the systematic review of studies done in a particular area, and has the strongest clinical evidence, being at the top of the hierarchy of evidence [4]. In order to improve meta-analysis method-ologies, we propose applying a generalized Box-Cox method to non-normal data to ac-count for negative input data.

## 2. Materials and Methods

In this paper, we focus on a few sample summary statistics, using the following notation: minimum value ($Q_{min}$), first quartile ($Q_1$), median ($Q_2$), third quartile ($Q_3$), maximum value ($Q_{max}$), mean ($\bar{x}$), standard deviation (s), and sample size (n). We will use $\hat{x}$ and $\hat{s}$ to denote the estimated sample mean and standard deviation, respectively. We will consider the three scenarios of reported summary statistics presented by Wan, Luo, and McGrath:

$$S_1: \{Q_{min}, Q_2, Q_{max}\}$$

$$S_2: \{Q_1, Q_2, Q_3\}$$

$$S_3: \{Q_{min}, Q_1, Q_2, Q_3, Q_{max}\}$$

**Comparator Methods**

Luo et al.'s method outperformed its predecessors for estimating sample mean. One key component introduced in this method is weighting the quantiles based on sample size. The weights allow the function to minimize the mean squared error of the estimator. For each of the scenarios, Luo proposed the following methods:

$$\hat{x} = \left(\frac{4}{4+n^{0.75}}\right)\frac{Q_{min}+Q_{max}}{2} + \left(\frac{n^{0.75}}{4+n^{0.75}}\right)Q_2 \quad \text{in } S_1$$

$$\hat{x} = \left(0.7 + \frac{0.39}{n}\right)\frac{Q_1+Q_3}{2} + \left(0.3 - \frac{0.39}{n}\right)Q_2 \quad \text{in } S_2$$

$$\hat{x} = \left(\frac{2.2}{2.2+n^{0.75}}\right)\frac{Q_{min}+Q_{max}}{2} + \left(0.7 - \frac{0.72}{n^{0.55}}\right)\frac{Q_1+Q_3}{2} + \left(0.3 + \frac{0.72}{n^{0.55}} - \frac{2.2}{2.2+n^{0.75}}\right)Q_2 \quad \text{in } S_3$$

Similarly, Wan et al. proposed an estimator for the standard deviation under each of the three scenarios:

$$\hat{s} = \frac{Q_{max}-Q_{min}}{2\varphi^{-1}(\frac{n-0.375}{n+0.25})} \quad \text{in } S_1$$

$$\hat{s} = \frac{Q_3-Q_1}{2\varphi^{-1}(\frac{0.75n-0.125}{n+0.25})} \quad \text{in } S_2$$

$$\hat{s} = \frac{Q_{max}-Q_{min}}{4\varphi^{-1}(\frac{n-0.375}{n+0.25})} + \frac{Q_3-Q_1}{4\varphi^{-1}(\frac{0.75n-0.125}{n+0.25})} \quad \text{in } S_3$$

These estimators are derived through the contrast of the distribution standard deviation and the expected order statistic values under a normal distribution.

In order to account for the data that do not follow a normal distribution, McGrath et al. proposed a new estimator for sample mean and standard deviation. The method builds on the previous Luo and Wan (Luo/Wan) methods but introduces a Box-Cox transformation to fit the quantiles into a normal distribution. Box-Cox equations are as follows, with $f_\lambda$ denoting the BC transformation and $f_\lambda(x)$ denoting the transformed quantiles:

$$f_\lambda(x_i) = y_i = \begin{cases} \frac{x_i^\lambda - 1}{\lambda} & \text{if } \lambda \neq 0 \\ ln(x_i) & \text{if } \lambda = 0 \end{cases}$$

Similarly, the inverse BC transformations, denoted by $f_\lambda^{-1}$ are as follows:

$$f_\lambda^{-1}(y_i) = x_i = \begin{cases} (\lambda * y_i + 1)^{1/\lambda} & \text{if } \lambda \neq 0 \\ exp(y_i) & \text{if } \lambda = 0 \end{cases}$$

In the McGrath method, the input data is used to find an optimal lambda, which increases the likelihood of transforming the data to follow a normal distribution. The optimization equations for $\lambda$ are as follows:

$$f_\lambda(Q_{max}) - f_\lambda(Q_2) = f_\lambda(Q_2) - f_\lambda(Q_{min}) \quad \text{in } S_1$$
$$f_\lambda(Q_3) - f_\lambda(Q_2) = f_\lambda(Q_2) - f_\lambda(Q_1) \quad \text{in } S_2$$
$$\underset{\lambda}{argmin}[(f_\lambda(Q_3) - f_\lambda(Q_2) - (f_\lambda(Q_2) - f_\lambda(Q_1)))^2 + (f_\lambda(Q_{max}) - f_\lambda(Q_2) - (f_\lambda(Q_2) - f_\lambda(Q_{min})))^2]$$
$$\text{in } S_3$$

The above formulas optimize lambda by finding a value such that the distance between the minimum and maximum values (in $S_1$), first and third quantiles (in $S_2$), or utilizes all quantiles ($S_3$) to make the distance from the left and right side of the median equidistant from each other.

Thus, the values are first transformed to follow a normal distribution using the optimized lambda, then the Luo/Wan methods are applied to the transformed values, and inverse transformed to get the final estimated mean and standard deviation.

**Proposed Method**

Since the Box-Cox method does not take negative values, we implement a generalized BC method, proposed by Yeo-Johnson et al. [5] In addition to implementing the generalized Box-Cox method for our method, we also propose using a maximum likelihood estimator (MLE) to optimize lambda, as compared to the calculation-based method proposed in McGrath.

The generalized Box-Cox equations are as follows:

When x ≥ 0, $f_\lambda(x_i) = y_i = \begin{cases} \frac{(x_i+1)^\lambda - 1}{\lambda} & if\ \lambda \neq 0 \\ ln(x_i + 1) & if\ \lambda = 0 \end{cases}$

When x < 0, $f_\lambda(x_i) = y_i = \begin{cases} \frac{-((-x_i+1)^{2-\lambda} - 1)}{2-\lambda} & if\ \lambda \neq 2 \\ -ln(-x_i + 1) & if\ \lambda = 2 \end{cases}$

As with the McGrath method, λ is an unknown value. We use the maximum likelihood estimator of lambda, which we define below,

$$\hat{\lambda}_{MLE} = \arg\min_\lambda \Pi_{i=1}^s Pr(f_\lambda(s_i)|\lambda),\ \text{where}\ Pr(a) = \phi_{\mu,\sigma^2}(a).$$

Once the generalized Box-Cox equations are applied, we use the Luo and Wan methods to calculate the estimated sample mean and standard deviation under the transformed distribution. We then inverse-transformed our estimated values to yield the final mean and standard deviation estimations under the original distribution.

The algorithm is as follows:

---

**Algorithm 1** Generalized Box-Cox Method.

    Apply generalized Box-Cox to input set $Q$ of values.

    Search for $\hat{\lambda}$ that maximizes objective function (likelihood, McGrath quantile functions)

    **for** $q_i$ of $Q$ **do**

        Apply $f_{\hat{\lambda}}$ (generalized Box-Cox) to $q_i$, obtaining $Q'$.

    **end for**

    Apply Luo and Wan methods to find $\hat{\mu}'$ and $\hat{\sigma}'$ under transformed distribution.
    Apply $f_{\hat{\lambda}}^{-1}$ (inverse generalized Box-Cox) to find $\hat{\mu}$ and $\hat{\sigma}$.

## 3. Results

*3.1. Simulation Setting*

We compared the methods under different simulation settings, including the normal, beta, and gamma distributions. Under each simulation, we set different sample sizes (n=10 to 500) in increments of 10 to simulate random numbers according to the distributions. Next, we extracted different sets of summary statistics (S1, S2, and S3) as presented by Wan, Luo, and McGrath. The estimated mean and standard deviation by different methods were then compared with the sample mean and standard deviation using the average relative error (ARE). To increase the reliability of the results, we took an average of 50 simulations to create each point on the graph. Table 1. Summarized the simulation settings.

| Simulation | S1 | S2 | S3 |
|---|---|---|---|
| **Table 1.** Summary of results under different simulation settings | | | |
| Fig 1: Normal (mean = 100, standard deviation = 1) | Mean: McGrath and Generalized BC outperform Luo<br>SD: McGrath and Generalized BC outperform Luo | Mean: McGrath and Generalized BC outperform Luo<br>SD: McGrath and Generalized BC outperform Luo | Mean: All methods perform similarly<br>SD: All methods perform similarly |
| Fig 2: Normal (mean = -100, standard deviation = 20)<br><br>*MG method does not work | Mean: Generalized BC outperforms Luo<br>SD: Generalized BC outperforms Wan | Mean: Generalized BC outperforms Luo<br>SD: Generalized BC outperforms Wan | Mean: Generalized BC and Luo perform similarly perform similarly<br>SD: Wan performs slightly better than Generalized BC |
| Fig 3: Beta (shape1 = 100, shape2 = 1) | Mean: McGrath and Generalized BC outperform Luo<br>SD: McGrath outperforms Generalized BC and Wan methods | Mean: McGrath and Generalized BC outperform Luo<br>SD: McGrath and Generalized BC outperform Wan | Mean: All perform similarly<br>SD: Generalized BC outperforms both Wan and McGrath |
| Fig 4: Negative Beta (shape1 = 100, shape2 = 1)<br><br>*MG method does not work | Mean: Generalized BC and Luo perform similarly<br>SD: Generalized BC outperforms Wan | Mean: Generalized BC and Luo perform similarly<br>SD: Generalized BC outperforms Wan | Mean: Generalized BC and Luo perform similarly<br>SD: Generalized BC and Wan perform similarly |
| Fig 5: Gamma (shape = 0.1, rate = 0.1) | Mean: McGrath outperforms Generalized BC and Luo<br>SD: McGrath outperforms Generalized BC and Wan methods | Mean: For the most part, Generalized BC performs the best<br>SD: For the most part, Generalized BC performs the best | Mean: McGrath and Luo outperform the Generalized BC method<br>SD: McGrath and Wan outperform Generalized BC |
| Fig 6: Negative Gamma (shape = 0.1, rate = 0.1)<br><br>*MG method does not work | Mean: Generalized BC and Luo perform similarly<br>SD: Wan slightly outperforms generalized BC | Mean: Generalized BC outperforms Luo<br>SD: Generalized BC outperforms Wan | Mean: Luo outperforms generalized BC<br>SD: Wan outperforms generalized BC |

*3.2. Results*

The McGrath (MG) method does not work when there are negative values in the data. Thus, we proposed applying the Generalized Box-Cox method to the sample mean and standard deviation estimations derived from Wan and Luo. We tested the values with different distributions, including normal (Figure 1), beta (Figure 2), and gamma (Figure 3). In Figure 1, we generated sample data to follow a normal distribution with a mean of 100 and a standard deviation of 1. We then applied all three scenarios to find $\hat{x}$ and $\hat{s}$, and compared these to the sample mean from the data we generated through the average relative error (ARE). We repeated the process with a beta distribution in Figure 2, with shape1 = 100 and shape2 = 1, and a gamma distribution in Figure 3, with shape = 0.1 and rate = 0.1. When the sample data is positive, the MG method and proposed method perform with comparable accuracy. Both Box-Cox methods outperformed the original Luo and Wan methods for calculating sample mean and standard deviation, respectively. When simulated under the normal distribution with a mean of 100 and standard deviation of 20, the McGrath method produced an error and thus is not included in Table 1.

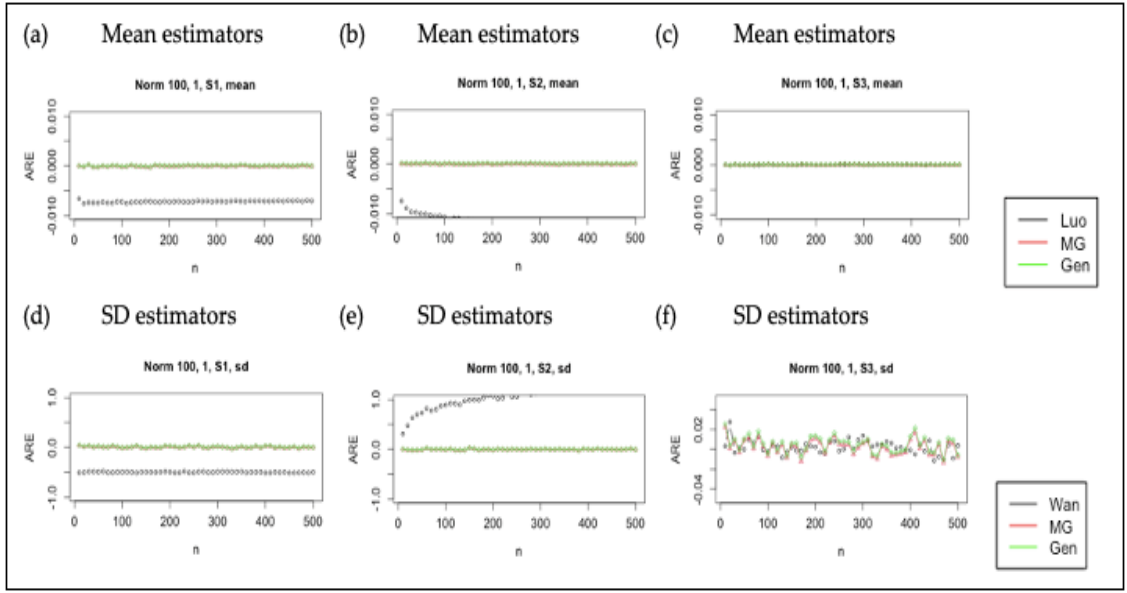

Figure 1. Results from a normal distribution (mean: 100 standard deviation: 1) comparing the Average Relative Error (ARE) among Luo/Wan(black), McGrath(red), and Generalized BC (green) for sample mean and standard deviation.

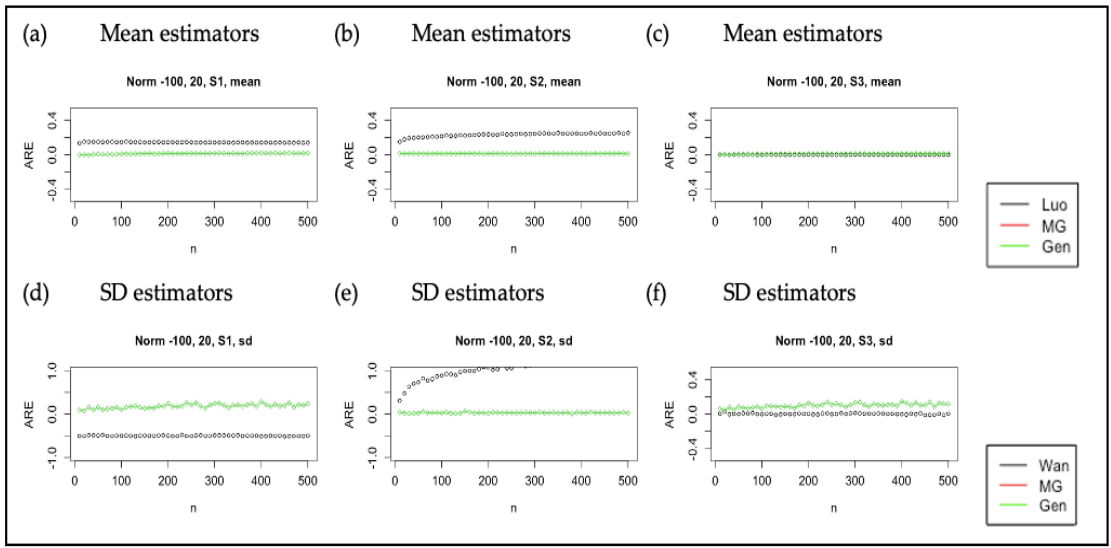

Figure 2. Results from a normal distribution (mean: -100 standard deviation: 20) comparing the Average Relative Error (ARE) among Luo/Wan(black), McGrath(red), and Generalized BC (green) for sample mean and standard deviation.

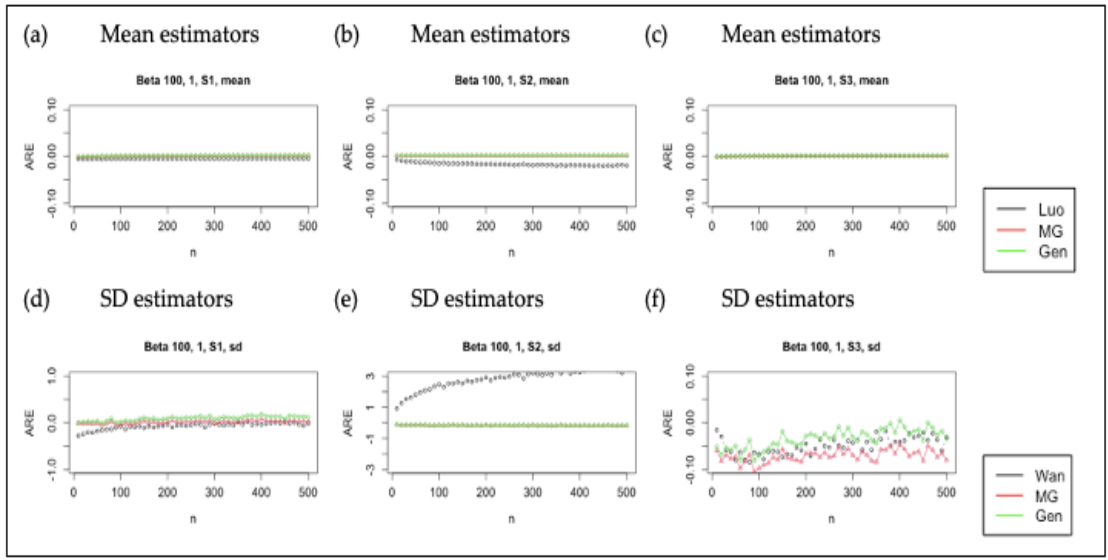

Figure 3. Results from a beta distribution (shape1: 100 shape2: 1) comparing the Average Relative Error (ARE) among Luo/Wan(black), McGrath(red), and Generalized BC (green) for sample mean and standard deviation.

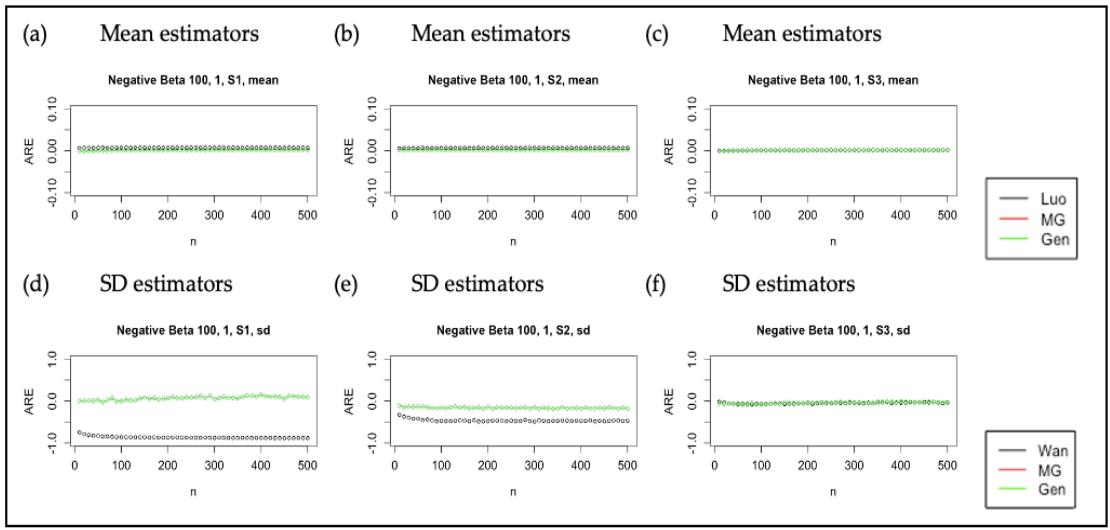

Figure 4. Results from a negative beta distribution (shape1: 100 shape2: 1) comparing the Average Relative Error (ARE) among Luo/Wan(black), McGrath(red), and Generalized BC (green) for sample mean and standard deviation.

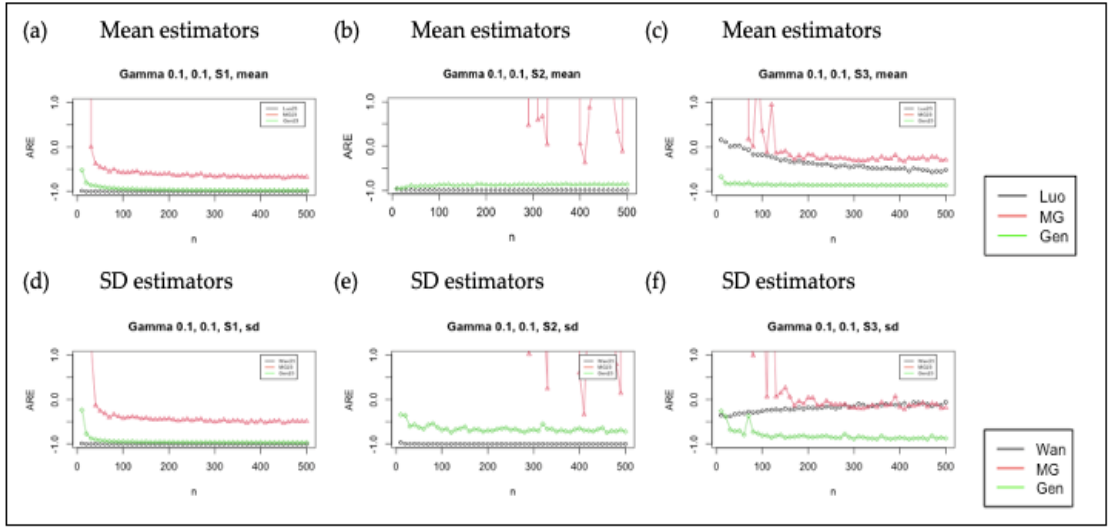

Figure 5. Results from a gamma distribution (shape: 0.1 rate: 0.1) comparing the Average Relative Error (ARE) among Luo/Wan(black), McGrath(red), and Generalized BC (green) for sample mean and standard deviation.

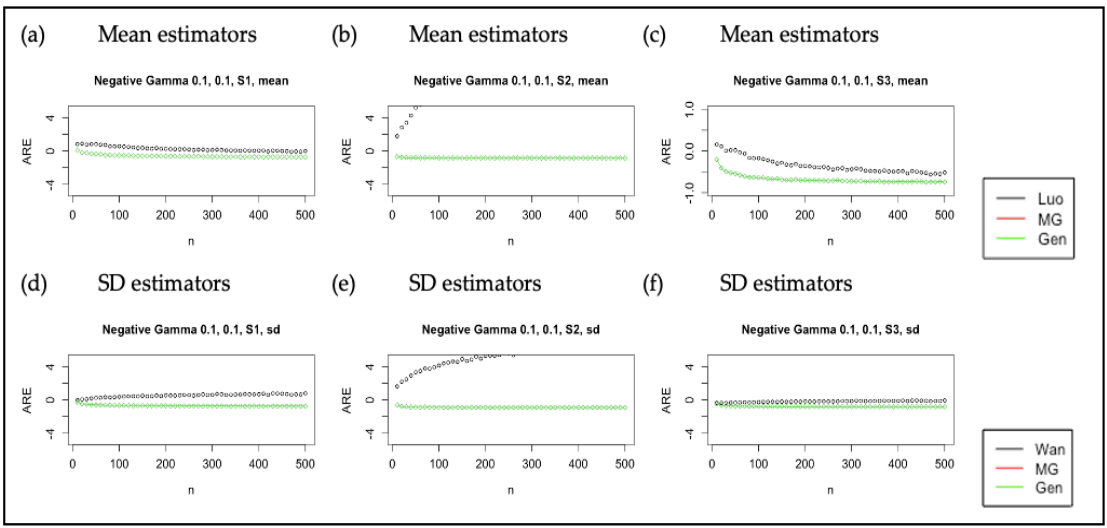

Figure 6. Results from a negative gamma distribution (shape: 0.1 rate: 0.1) comparing the Average Relative Error (ARE) among Luo/Wan(black), McGrath(red), and Generalized BC (green) for sample mean and standard deviation.

## 4. Discussion

In this paper, we outlined a new method for estimating sample mean and standard deviation. This method is especially helpful when encountering non-normal data with negative values. In the future, we hope to improve this method further by narrowing down the conditions under which this method works best and testing different maximum likelihood estimators.

One possible method to avoid encountering negative value error with the MG box-cox method is to introduce a rightward shift. However, upon trial, this method still returns error in some cases and is therefore not ideal compared to using the generalized box-cox method.

To conclude, our method works best for dealing with data that includes positive and negative values. We also hope to create an R package for public access to our method and include real-world example data in medicine and education.

**Supplementary Materials:** N/A

**Author Contributions:** Conceptualization, O.X. and S.W.; methodology, O.X. and S.W; validation, O.X. and S.W.; formal analysis, O.X. and S.W.; investigation, O.X. and S.W.; writing—original draft preparation, O.X., S.W. and M.C; writing—review and editing, O.X., S.W. and M.C.; visualization, O.X.; supervision, S.W. and M.C. All authors have read and agreed to the published version of the manuscript.

**Funding:** This research received no external funding

**Conflicts of Interest:** The authors declare no conflict of interest.

**References**
1.  Luo et al., Optimally estimating the sample mean from the sample size, median, mid-range, and/or mid-quartile range; *Statistical Methods in Medical Research* in 2016
2.  Wan et al., Estimating the sample mean and standard deviation from the sample size, median, range and/or interquartile range*; BMC Medical Research Methodology* in 2014
3.  McGrath et al., Estimating the sample mean and standard deviation from commonly reported quantiles in meta-analysis; *Stat Methods Med Res.* in 2021Haidich, Meta-analysis in medical research; *Hippokratia* in 2010
4.  Yeo and Johnson, A new family of power transformations to improve normality or symmetry; *Biometrika* in 2000